\documentclass{aa}  

\usepackage{graphicx}
\usepackage{txfonts}
\usepackage{booktabs}
\usepackage{xcolor}
\usepackage{threeparttable}
\usepackage[colorlinks=True, citecolor=blue, linkcolor=blue, urlcolor=blue]{hyperref}
\begin{document}

\title{A 100 kpc Ram Pressure Tail Trailing the Group Galaxy NGC 2276}

\author{I.D. Roberts\inst{\ref{leiden},\ref{waterloo},\ref{wca}}
        \and
        R.J. van Weeren\inst{\ref{leiden}}
        \and
        F. de Gasperin\inst{\ref{inaf_bologna}}
        \and
        A. Botteon\inst{\ref{inaf_bologna}}
        \and
        H.W. Edler\inst{\ref{hamburg}}
        \and
        A. Ignesti\inst{\ref{inaf_padova}}
        \and
        L. Matijevi\'{c}\inst{\ref{zagreb}}
        \and
        N. Tomi\v{c}i\'{c}\inst{\ref{zagreb}}
        }

\institute{
    Leiden Observatory, Leiden University, PO Box 9513, 2300 RA Leiden, The Netherlands \label{leiden}
    \and
    Department of Physics \& Astronomy, University of Waterloo, Waterloo, ON N2L 3G1, Canada
    \label{waterloo}
    \and
    Waterloo Centre for Astrophysics, University of Waterloo, 200 University Ave W, Waterloo, ON N2L 3G1, Canada
    \label{wca}
    \and
    INAF-Istituto di Radioastronomia, Via Gobetti 101, I-40129 Bologna, Italy
    \label{inaf_bologna}
    \and
    Hamburger Sternwarte, University of Hamburg, Gojenbergsweg 112, 21029, Hamburg, Germany
    \label{hamburg}
    \and
    INAF--Padova Astronomical Observatory, Vicolo dell'Osservatorio 5, I-35122 Padova, Italy \label{inaf_padova}
    \and
    Department of Physics, Faculty of Science, University of Zagreb, Bijeni\v{c}ka Cesta 32, 10000, Zagreb, Croatia \label{zagreb}
    }

% \abstract{}{}{}{}{} 
% 5 {} token are mandatory
 
  \abstract{We present the discovery of a $100\,\mathrm{kpc}$ low-frequency radio tail behind the nearby group galaxy, NGC 2276. The extent of this tail is a factor of ten larger than previously reported from higher-frequency radio and X-ray imaging. The radio morphology of the galaxy disc and tail suggest that the tail was produced via ram-pressure stripping, cementing NGC 2276 as the clearest known example of ram-pressure stripping in a low-mass group. With multi-frequency imaging, we extract radio continuum spectra between ${\sim}50\,\mathrm{MHz}$ and $1.2\,\mathrm{GHz}$ as a function of projected distance along the tail. All of the spectra are well fit by a simple model of spectral ageing due to synchrotron and inverse-Compton losses. From these fits we estimate a velocity of $870\,\mathrm{km\,s^{-1}}$ for the stripped plasma across the plane of the sky, and a three-dimensional orbital velocity of $970\,\mathrm{km\,s^{-1}}$ for NGC 2276. The orbital speed that we derive is in excellent agreement with the previous estimate from Rasmussen et al., despite it being derived with a completely independent methodology.}


   \maketitle
%
%-------------------------------------------------------------------

\section{Introduction} \label{sec:intro}

The NGC 2300 galaxy group is a well-studied system for two, likely interconnected, reasons. It was the first low-mass group ($M_\mathrm{200} \simeq 3\times10^{13}\,\mathrm{M_\odot}$), outside of compact groups, to have detected diffuse X-ray emission from the intra-group medium \citep{mulchaey1993}. Secondly, it hosts the star-bursting satellite galaxy NGC 2276, which was among the first known examples of ongoing ram-pressure stripping in a group environment \citep[e.g.][]{condon1983,mulchaey1993}.
\par
A peculiar optical morphology for NGC 2276 was noted by \citet{arp1966} and further detailed by \citet{shakhbazyan1973} who identified a number of compact star-forming regions confined to the western edge of the disc. In addition, compressed radio continuum emission on the western edge and a short radio tail to the east were observed at $1.4\,\mathrm{GHz}$ by \citet{condon1983}. This suggests a scenario where NGC 2276 is moving from east-to-west across the plane of the sky leading to ram-pressure-induced star formation on the western edge and the formation of a tail from the stripped interstellar medium (ISM) to the east. This framework is supported by more recent observations that have identified an accompanying X-ray tail the to the east of NGC 2276 \citep{rasmussen2006}, increased star formation efficiency and an unusually large number of ultraluminous X-ray sources along the western edge \citep{wolter2015,tomicic2018,fadda2023}, as well enhanced \textsc{[Cii]} emission due to shocks from ram pressure \citep{fadda2023}. Thus NGC 2276 is a clear example of a nearby star-forming galaxy [$\mathrm{SFR} \simeq 10\,\mathrm{M_\odot\,yr^{-1}}$, \citealt{wolter2015} (X-ray emission), \citealt{tomicic2018} ($\mathrm{H\alpha}$ and FUV+$22\,\mathrm{\mu m}$)] experiencing ram-pressure stripping, and as such is an excellent laboratory in order to probe the impact of the more modest group environment on star formation quenching.
\par
This becomes relevant in the broader context as the majority of galaxies in the low-$z$ Universe are found in galaxy groups, far outnumbering galaxies in massive clusters \citep{geller1983,eke2005,robotham2011}. This, combined with the fact that the quenched fraction for group galaxies is significantly larger than that for field galaxies \citep[e.g.][]{wetzel2012}, demonstrates that a substantial amount of environmental quenching must occur in these lower-mass groups. The importance of understanding environmental quenching in groups is furthered when one considers that a significant fraction of galaxies on the cluster red sequence at $z=0$ were not quenched in a cluster-mass halo, but instead were `pre-processed' in a lower mass group prior to cluster infall \citep[e.g.][]{mcgee2009,delucia2012,bahe2013,roberts2017,jung2018,oxland2024}.
\par
In galaxy clusters, some consensus is emerging on environmental quenching being primarily driven by ram-pressure stripping \citep[e.g.][]{vollmer2001,boselli2006,brown2017,poggianti2017,roberts2019,roberts2021_LOFARclust,werle2022}. Either by directly removing the atomic and/or molecular ISM, or by stripping the circumgalactic medium and therefore removing the supply for future gas condensation onto the disc (often referred to as `starvation', see \citealt{cortese2021,boselli2022_review} for recent reviews). Given the lower X-ray luminosities (tracing the intra-group/cluster medium) and velocity dispersions in groups relative to clusters, ram pressure (which scales as $\rho_\mathrm{ICM}\,v^2$) will be weaker in these lower-mass environments, and thus it is less clear whether it is strong enough to serve as the primary quenching mechanism. \citet{roberts2021_LOFARclust} find a number of radio continuum `jellyfish galaxies' (with one-sided radio tails likely due to ram-pressure stripping) in low-$z$ clusters. The long radio-continuum tails observed in cluster galaxies implies that the stripped ISM transports non-thermal components, such as magnetic field and cosmic rays, which lose energy via synchrotron radiation. These stripped cosmic rays originate from supernova shock acceleration in the galaxy disc, and are subsequently transported along the tail via ram pressure. Thus the radio emission in the tail is tracing past star formation from the galaxy, not current star formation in the tail \citep[e.g.][]{roberts2024_radiospec_coma}. \citet{roberts2021_LOFARgrp} show that similar galaxies are also found in groups, though at a reduced frequency relative to clusters -- especially for low-mass groups comparable to the NGC 2300 system. Thus, NGC 2276 is a rare example of clear ram-pressure stripping in a low-mass group and therefore studying this object represents a unique opportunity to constrain the process in a sparse environment.
\par
This work is the first of a series of papers using new multi-frequency observations of NGC 2276 (X-ray, UV, optical imaging and spectroscopy, sub-mm, and low-frequency radio) to constrain the impact from environment on star formation in this extreme galaxy. In this work we present new radio continuum observations across a range of low frequencies ($50\,\mathrm{MHz} - 1.2\,\mathrm{GHz}$). Given the vicinity of NGC 2276 ($D \sim 30\,\mathrm{Mpc}$), we reach very low surface brightness and sub-kiloparsec spatial resolution, allowing us to map in detail the radio continuum emission and spectral index across the galaxy disc and stripped tail. This in turn constrains the speed at which material is being removed from the disc as well as the three-dimensional speed of NGC 2276 relative to the intragroup medium. We note that this work presents the lowest frequency observations ever of a ram-pressure stripped tail (${\sim}50\,\mathrm{MHz}$) -- either in a group or a cluster.
\par
The outline of this paper is as follows. In Sect.~\ref{sec:data} we describe the observations and reduction of the multi-frequency radio data used in this work. In Sect.~\ref{sec:radio_images} we present our new multi-frequency radio continuum imaging of NGC 2276 and its stripped tail, including spectral index maps. In Sect.~\ref{sec:spec_age} we extract radio continuum spectra in slices along the tail. We then fit those spectra with a synchrotron ageing model in order to infer plasma ages and estimate the speed at which material is transported down the tail. Finally, in Sects. \ref{sec:discussion} \& \ref{sec:conclusion} we discuss and summarize the main results from this work. Throughout the paper we assume $H_0 = 75\,\mathrm{km\,s^{-1}\,Mpc^{-1}}$ and a distance to NGC 2276 of $36.8\,\mathrm{Mpc}$ \citep{tully1988}. At this distance $1\arcsec \simeq 180\,\mathrm{pc}$. The radio synchrotron spectrum is described as $S_\nu \propto \nu^\alpha$, where $S_\nu$ is the flux density, $\nu$ is the frequency, and $\alpha$ is the spectral index.

\section{Data} \label{sec:data}

\begin{table*}[!ht]
\begin{threeparttable}
\centering
\caption{Radio image parameters for low-resolution and high-resolution (in parentheses) images.}
\label{tab:sensitivity_table}
\begin{tabular}{l c c c c c c c c c}
\toprule
\toprule
Band & $t_\mathrm{source}$ & $\nu$ & $\Delta \nu$ & Robust & $b_\mathrm{maj}$ & $b_\mathrm{min}$ & PA & rms & $\delta_\mathrm{cal}$\tnote{2} \\
& h & MHz & MHz & & $\arcsec$ & $\arcsec$ & $\deg$ & $\mathrm{\mu Jy \, bm^{-1}}$ & \\
\midrule
LOFAR LBA & 9.0\tnote{1} & 54 & 24 & 0 (-1) & $24.2\;(11.8)$ & $20.0\;(10.7)$ & 127 (73) & $2.2\;(2.9) \times 10^3$ & 0.10 \\
LOFAR HBA & 8.0 & 144 & 48 & 0 (-1) & $11.9\;(4.9)$ & $8.2\;(4.3)$ & 125 (106) & $5.9\;(8.0) \times 10^1$ & 0.10 \\
uGMRT B3 & 3.9 & 400 & 200 & 0.5 (-1) & $17.8\;(11.3)$ & $7.4\;(3.7)$ & 22 (27) & $3.9\;(3.1) \times 10^1$ & 0.05 \\
uGMRT B4 & 7.5 & 650 & 200 & 0.25 (-1) & $8.0\;(3.9)$ & $4.8\;(2.5)$ & 37 (38) & $0.8\;(1.1) \times 10^1$ & 0.05 \\
uGMRT B4 low & 7.5 & 600 & 100 & 0.25 (-1) & $8.8\;(4.2)$ & $5.2\;(2.7)$ & 41 (40) & $1.0\;(1.5) \times 10^1$ & 0.05 \\
uGMRT B4 high & 7.5 & 700 & 100 & 0.25 (-1) & $7.4\;(3.7)$ & $4.4\;(2.3)$ & 33 (35) & $0.9\;(1.6) \times 10^1$ & 0.05 \\
uGMRT B5 & 4.8 & 1230 & 333 & 0.25 (-0.5) & $6.9\;(4.6)$ & $4.2\;(2.7)$ & 113 (112) & $2.5\;(3.8) \times 10^1$ & 0.05 \\
\bottomrule
\end{tabular}
\begin{tablenotes}
    \item [1] This is the combined on-source time for the three LoLSS pointings covering NGC 2276. For all three, NGC 2276 is offset from pointing centre and therefore the effective on-source time, factoring in the diminishing primary-beam response, is certainly $<\!9\,\mathrm{h}$.

    \item [2] The relative calibration uncertainty on the flux-scale.
\end{tablenotes}
\end{threeparttable}
\end{table*}

This work makes use of multi-frequency radio continuum imaging from the Low Frequency Array (LOFAR, \citealt{vanhaarlem2013}) and the upgraded Giant Metrewave Telescope (uGMRT). Below we briefly describe the data calibration and imaging procedures for each of the frequency bands.

\subsection{LOFAR} \label{sec:data_lofar}

\subsubsection{High-band Antenna} \label{data:hba}

The LOFAR $144\,\mathrm{MHz}$ high-band antenna (HBA) data used in this work were taken from the LOFAR Two-metre Sky Survey (LoTSS, \citealt{shimwell2019,shimwell2022}). At a declination of $+85\deg$, the LoTSS data for NGC 2276 are not part of the three LoTSS public data releases to date. That said, observations of the NGC 2276 field were already taken and processed with the standard LoTSS pipelines. Full details on the observation, calibration, and imaging strategy for LoTSS are given in \citet{shimwell2019,tasse2021,shimwell2022}. These data will be made publicly available in a future LoTSS data release.
\par
NGC 2276 falls within the primary beam of LoTSS pointing P113+87 at an offset of $1.13\deg$ from the pointing centre. As is standard for LoTSS, P113+87 was observed for an $8\,\mathrm{h}$ integration time. Its prominent radio continuum tail is clearly visible in the LoTSS pipeline-processed image; however, in order to maximize image quality we extracted and reprocessed a $30\arcmin \times 30\arcmin$ region centred on NGC 2276 from the larger LoTSS image. This extraction and self-calibration procedure for LOFAR targets is described in detail in \citet{vanweeren2021}. Lastly, we apply a factor of 1.7 in order to align the dataset with the NVSS flux-scale. The method used to derive this flux-scale correction factor is described in \citet{shimwell2022}.

\subsubsection{Low-band Antenna} \label{data:lba}

The data for NGC 2276 at $54\,\mathrm{MHz}$ is taken from the wide-area survey for the LOFAR low-band antenna (LBA), the LOFAR LBA Sky Survey (LoLSS, \citealt{degasperin2021,degasperin2023}). Similar to the HBA, the LBA data for NGC 2276 are not part of the current public data releases for LoLSS \citep{degasperin2021,degasperin2023}, but will be made public in a future date release.
\par
NGC 2276 falls within the primary beam of three distinct LoLSS pointings: P098+84, P113+87, and P125+85, each observed for a $3\,\mathrm{h}$ integration.  Each pointing was processed individually with the Pipeline for LOFAR LBA (PiLL\footnote{\url{https://github.com/revoltek/LiLF}}, \citealt{degasperin2023}). We extracted and self-calibrated a $50\arcmin \times 50\arcmin$ region around NGC 2276 for each of the LBA pointings, again following the methodology of \citet{vanweeren2021}.
\par
For both the HBA and the LBA, the largest recoverable angular scale (LAS) is $>\!1\deg$, far larger than the source extent of NGC 2276 and its stripped tail (${\sim}10\arcmin \times 2\arcmin$, see Fig.~\ref{fig:radio_imgs}). Thus we are confident that we are not resolving out any flux at these frequencies.

\subsection{uGMRT} \label{sec:data_ugmrt}

\begin{figure}
    \centering
    \includegraphics[width = \columnwidth]{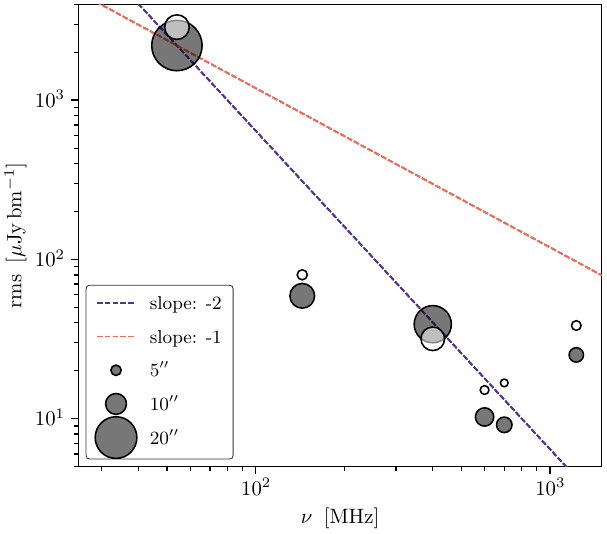}
    \caption{Noise in our radio maps as a function of frequency for the radio continuum images used in this work. The grey markers correspond to the low-resolution images and the white markers correspond to the high resolution images. The marker size scales with the size of the major axis of the restoring beam. For reference, lines corresponding to spectral indices of -1 and -2 are shown through our lowest frequency band.}
    \label{fig:rms_plot}
\end{figure}

The uGMRT data for bands three ($300 - 500\,\mathrm{MHz}$), four ($550 - 750\,\mathrm{MHz}$), and five ($1020 - 1420\,\mathrm{MHz}$) were obtained on 2023-11-25 (observation no.\ 16101), 2023-11-26 (observation no.\ 16106), and 2023-11-27 (observation no.\ 16111), respectively (P.I.\ Roberts). On-source times for bands three, four, and five were $3.9\,\mathrm{h}$, $7.5\,\mathrm{h}$, and $4.8\,\mathrm{h}$, respectively, bracketed by ${\sim}10\,\mathrm{min}$ flux calibrator observations.
\par
The data were downloaded from the GMRT online archive\footnote{\url{https://naps.ncra.tifr.res.in/goa/data/search}} and calibrated using the Source Peeling and Atmospheric Modeling pipeline (\texttt{SPAM}, \citealt{intema2009}), following the wide-band GMRT workflow\footnote{\url{https://www.intema.nl/doku.php?id=huibintemaspampipeline\#experimentalprocessing\_ugmrt\_wideband_data}}. The band three observation was split into six $33\,\mathrm{MHz}$ sub-bands, the band four into four $50\,\mathrm{MHz}$ sub-bands, and the band five into six $66\,\mathrm{MHz}$ sub-bands.  Each of these sub-bands were then processed with the main \texttt{SPAM} pipeline. We note that for band five we excluded the highest frequency sub-band as it contains the \textsc{H$\,$i} $21\,\mathrm{cm}$ line. We leave an analysis of the atomic hydrogen content of NGC 2276 for a future work. Given that this sub-band is at the edge of the bandwidth response function, this exclusion did not strongly impact the final RMS noise that we reached for the band five image.
\par
The LAS that the uGMRT is sensitive to is ${\sim}30\arcmin$ for band three, ${\sim}20\arcmin$ for band four, and ${\sim}10\arcmin$ for band five. Comparing this to the ${\sim}10\arcmin \times 2\arcmin$ angular size of NGC 2276 and its stripped tail at LOFAR frequencies, we should be sensitive to emission from the full tail in both bands three and four. For band five the theoretical LAS is roughly the same size as the largest source dimension, though in practice the LAS is likely somewhat smaller. Thus we may be missing some large-scale flux in band five; however, we also note that due to the shorter synchrotron lifetimes of cosmic-ray electrons at higher frequencies, we expect that the tail will be intrinsically shorter at $\sim$gigahertz frequencies than $\sim$megahertz frequencies. Roughly speaking, for a simple model of synchrotron ageing, we expect that the tail will be a factor of three shorter at $1.2\,\mathrm{GHz}$ than $144\,\mathrm{MHz}$ \citep{ignesti2022_meerkat,roberts2024_radiospec_coma}, which would bring it within the LAS of band 5.

\subsection{Imaging} \label{sec:imaging}

All calibrated measurement sets are imaged with \texttt{WSCLEAN} \citep{offringa2014,offringa2017}. For the LBA data we image together the three extracted and re-calibrated LoLSS pointings. For the uGMRT, we image together the measurement sets for each of the sub-bands to produce a full-band image for each of the observing bands. For band four, given the high $\mathrm{S/N}$ of this dataset, in addition to the full-band image we also split the wide-band into two images (one centred on $600\,\mathrm{MHz}$, "uGMRT B4 low", and one centred on $700\,\mathrm{MHz}$, "uGMRT B4 high").  We use the full-band image for the spectral index maps shown in Sect.~\ref{sec:low-res} and the split-band images for the rest of the paper.
\par
For all frequency bands we produce both a low- and high-resolution image. The low-resolution image is optimized for analysis of the extended, diffuse emission in the stripped tail and the high-resolution image is used to optimize the spatial resolution for analysis of the galaxy disc. The details of all images, including imaging parameters, synthesized beam shapes, and achieved sensitivity, are listed in Table~\ref{tab:sensitivity_table} and the RMS noise as a function of observing frequency is shown in Fig.~\ref{fig:rms_plot}.

\section{Radio Images} \label{sec:radio_images}

\begin{figure*}
    \centering
    \includegraphics[width = \textwidth]{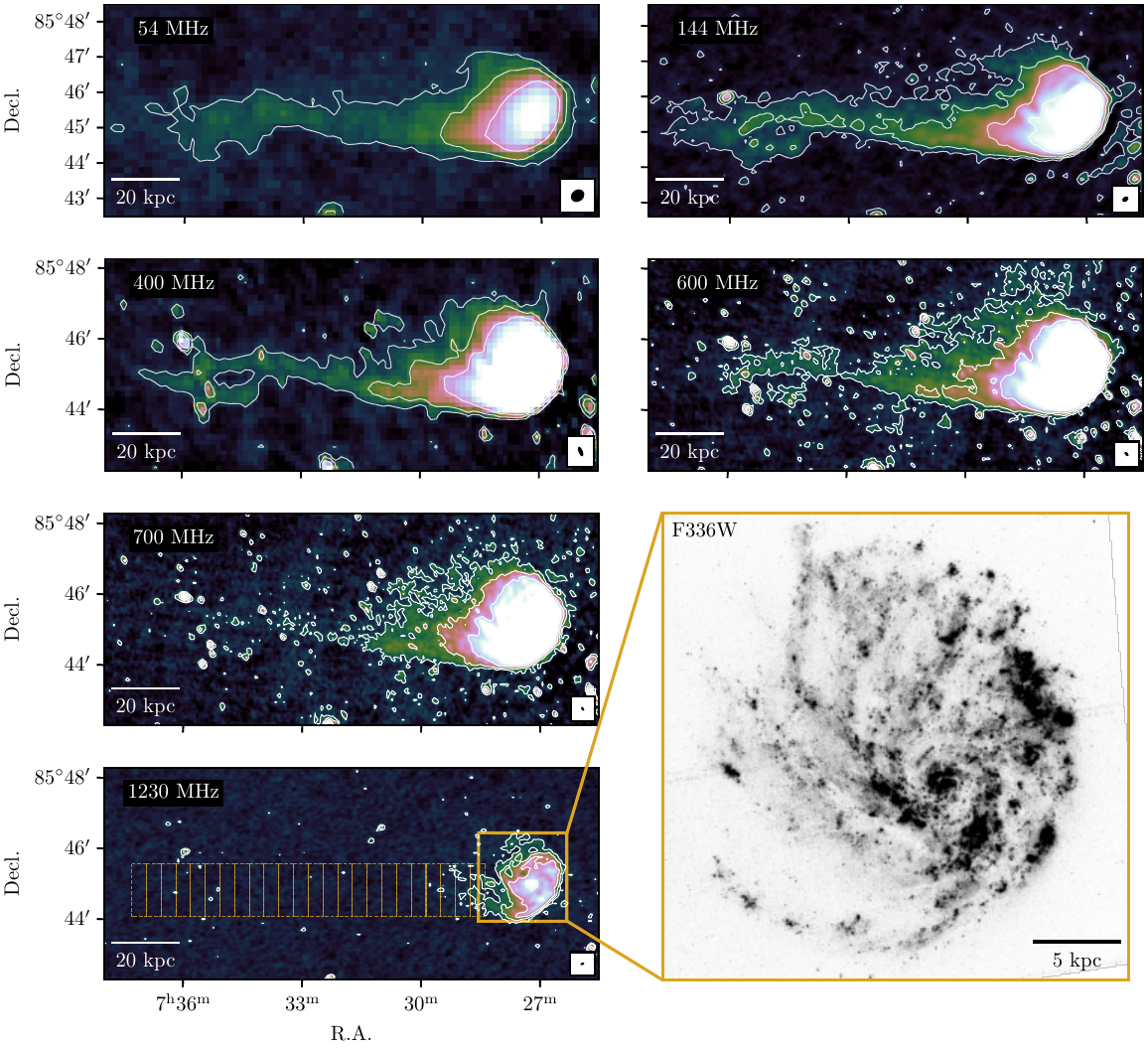}
    \caption{Radio images used in this work, increasing in frequency from left to right and top to bottom. Images are shown with an Asinh stretch set to highlight the extended tail, detailed images of the galaxy disc are shown in Fig.~\ref{fig:radio_imgs_highres}. Contour levels begin at $3 \times \mathrm{rms}$ and increase by factors of three. Each panel shows the restoring beam shape in the bottom-right corner (see Table~\ref{tab:sensitivity_table} for specfic values). For reference, we also show a near-UV image of the galaxy disc from the \textit{Hubble Space Telescope} (F336W, PI Sell), highlighting the strong star formation on the western edge of the disc. The apertures shown in the bottom-left panel are used to extract radio spectra along the tail (see Sect.~\ref{sec:spec_age})}
    \label{fig:radio_imgs}
\end{figure*}

\subsection{Low-resolution Images} \label{sec:low-res}

In Fig.~\ref{fig:radio_imgs} we plot the low-resolution radio continuum images of NGC 2276 ranging from $54\,\mathrm{MHz}$ to $1230\,\mathrm{MHz}$. For reference, we also show a near-UV image of the disc of NGC 2276 from the \textit{Hubble Space Telescope} (F336W, P.I.\ Sell).
\par
Most striking in Fig.~\ref{fig:radio_imgs} is the ${\sim}100\,\mathrm{kpc}$ (in projection) tail extending to the east of NGC 2276. This emission is tracing synchrotron radiation from cosmic-ray electrons accelerated by supernovae in the disc and subsequently stripped off to the east. The tail is an order of magnitude longer than previously observed at GHz and X-ray frequencies \citep{condon1983, rasmussen2006}. The tail is most prominent at the lowest frequencies and becomes progressively shorter towards higher frequencies. This is qualitatively consistent with the simple ageing of cosmic-ray electrons as they are transported along the tail -- a framework that we explore in quantitative detail in the subsequent sections.

\begin{figure*}
    \centering
    \includegraphics[width = \textwidth]{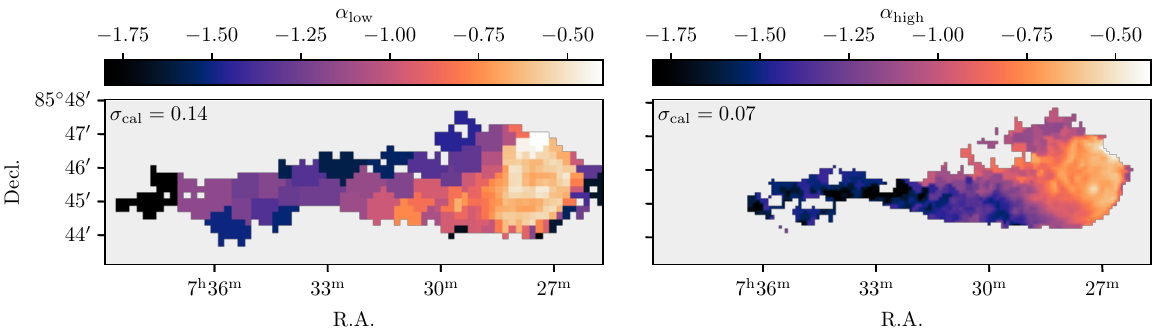}
    \caption{Spectral index maps at low- and high-frequency for NGC 2276 and its stripped tail. Both maps are Voronoi binned \citep{cappellari2003} to a target of $\mathrm{S/N} = 15$. In the upper left of each panel we list the uncertainty on the spectral index derived from the individual flux-calibration uncertainties (which is the dominant source of uncertainty given the high $\mathrm{S/N}$ from the Voronoi binning).}
    \label{fig:spec_maps_tail}
\end{figure*}

In Fig.~\ref{fig:spec_maps_tail} we show spectral index maps from the low-resolution products. We consider a low-frequency spectral index map, between $54\,\mathrm{MHz}$ and $144\,\mathrm{MHz}$, and a high-frequency spectral index map, between $144\,\mathrm{MHz}$ and $650\,\mathrm{MHz}$\footnote{Here we use the full $200\,\mathrm{MHz}$ bandwidth image from uGMRT band four.}. In both cases we smooth the higher resolution image to match the synthesized beam shape of the lower resolution image. To ensure uniform $\mathrm{S/N}$ across the spectral index maps we use Voronoi binning \citep{cappellari2003}. For the low-frequency spectral index maps we Voronoi bin according to the $54\,\mathrm{MHz}$ image (which is shallower than the $144\,\mathrm{MHz}$ image) and for the high-frequency spectral index map we bin according to the $650\,\mathrm{MHz}$ image. In both cases we set a target of $\mathrm{S/N} = 15$ for each bin and only include pixels which have $\mathrm{S/N} > 2$ in each of the two frequencies contributing to the spectral index map.
\par
The low- and high-frequency spectral index maps are shown in Fig.~\ref{fig:spec_maps_tail}. Given the relatively high $\mathrm{S/N}$ of the Voronoi bins, the uncertainty on the spectral indices is dominated by the flux-calibration uncertainties (see Table~\ref{tab:sensitivity_table}). For the low-frequency spectral index map this corresponds to an uncertainty of 0.14 and for the high-frequency map an uncertainty of 0.07.
\par
For the low-frequency spectral index map (Fig.~\ref{fig:spec_maps_tail}, left) the galaxy disc is highlighted by relatively flat emission ($-0.7 \lesssim \alpha \lesssim -0.5$) typical of recent acceleration from supernovae (the same is indeed true for $\alpha_\mathrm{high}$). We defer a comprehensive study of the radio emission and star formation within the disc of NGC 2276 to a future paper.  Broadly speaking, $\alpha_\mathrm{low}$ steepens with increasing distance from the galaxy disc, reaching $\alpha_\mathrm{low} < -1$ at the eastern terminus of the tail.
\par
For $\alpha_\mathrm{high}$, the spectral index also steepens along the tail reaching $\alpha_\mathrm{high} \lesssim -1.6$ at the tail end. This highlights the depth of the uGMRT band four data which allow us to probe these regions of ultra-steep spectral index. Comparing the maps for $\alpha_\mathrm{low}$ and $\alpha_\mathrm{high}$ (which are on the same colour scale), there appears to be positive curvature in the spectrum along the tail with $\alpha_\mathrm{low} > \alpha_\mathrm{high}$. Though given the uncertainties, roughly 0.14 and 0.07 on $\alpha_\mathrm{low}$ and $\alpha_\mathrm{high}$ respectively, this curvature is only marginally statistically significant at the ${\sim}2\sigma$ level.

\subsection{High-resolution Images} \label{sec:high-res}

\begin{figure} 
    \centering
    \includegraphics[width = \columnwidth]{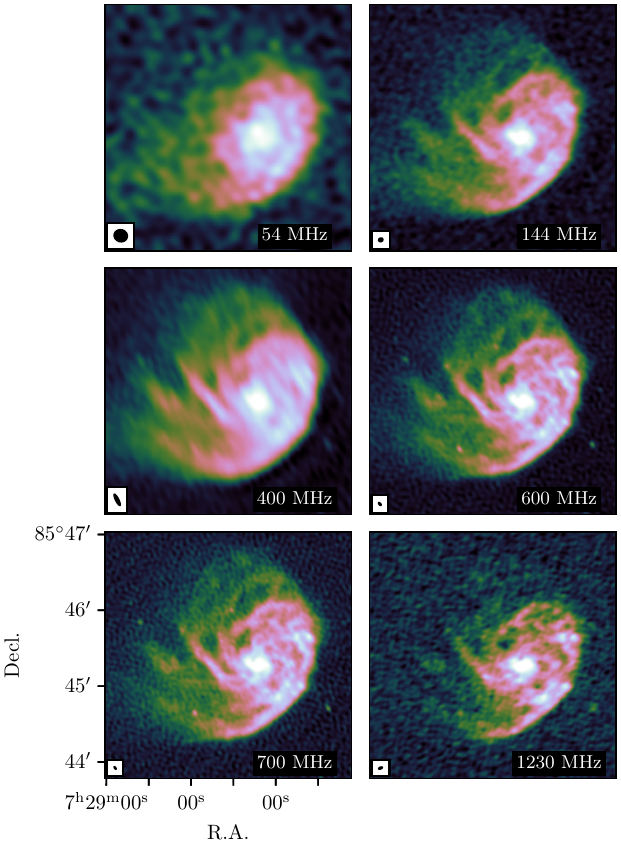}
    \caption{High-resolution radio images from $54$ to $1230\,\mathrm{MHz}$ showing the galaxy disc of NGC 2276. Images are displayed with an Asinh stretch. We show the FWHM beam shape in the lower left of each panel.}
    \label{fig:radio_imgs_highres}
\end{figure}

\begin{figure*}
    \centering
    \includegraphics[width = \textwidth]{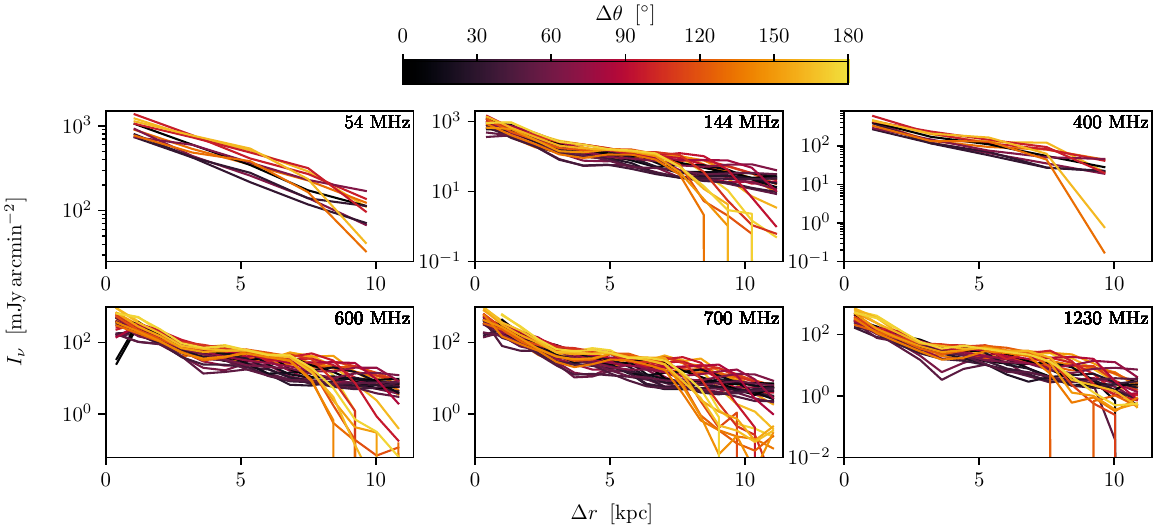}
    \caption{Radial surface brightness profiles as a function of azimuthal angle. Each panel corresponds to a different frequency band. Each solid line is a radial surface brightness profile measured within a wedge that is oriented at a specific azimuthal angle ($\Delta \theta$, see text for details).}
    \label{fig:galaxy_radio_profiles}
\end{figure*}

In Fig.~\ref{fig:radio_imgs_highres}, we show our high-resolution radio images of NGC 2276, zoomed in on the galaxy disc region. We only briefly make use of these high-resolution products in this paper, but they will be used extensively in future papers in the series focusing on star formation and cosmic-ray transport within the disc, where they will be combined with optical images (\textit{HST} and optical integral field spectra; \citealt{tomicic2018}, Matijevi\'{c} et al. in prep.). For the $144$, $600$, $700$, and $1230\,\mathrm{MHz}$ images we are able to achieve sub-kiloparsec resolution. At $400\,\mathrm{MHz}$ the resolution is limited by poor $u-v$ coverage due to the short observations, leading to a highly elliptical beam shape. For the LOFAR LBA at $54\,\mathrm{MHz}$, such high angular resolution would only be possible by incorporating the international LOFAR stations and by default LoLSS does not record data from the international stations.
\par
With the exception of the $54\,\mathrm{MHz}$ image, all high-resolution images show a sharp drop-off in flux at the western edge of the disc (opposite to the tail direction). The lack of the sharp edge in the $54\,\mathrm{MHz}$ image is likely related to the poor spatial resolution. We quantify the drop in flux density along the western edge in Fig.~\ref{fig:galaxy_radio_profiles} where we plot surface brightness profiles for each frequency measured as a function of azimuthal angle in wedges spanning between the galaxy centre and the edge of the disc. The size of the disc is given by $R_{25} = 1.1\arcmin$ (\href{http://atlas.obs-hp.fr/hyperleda/}{HyperLeda Database}). We set the opening angle of the wedges such that the arc length at $R = 0.5 \times R_{25}$ is equal to $1.5\times$ the FWHM of the beam major axis. This ensures that the pixels in neighbouring wedges are roughly independent and also means that the number of azimuthal wedges will vary with observing frequency. The $54\,\mathrm{MHz}$ image with the lowest spatial resolution will have coarser azimuthal resolution (11 wedges) compared to the $700\,\mathrm{MHz}$ image with the highest spatial resolution (36 wedges). For each azimuthal wedge, we make surface brightness profiles with radial steps equal to $1\times$ the FWHM of the beam major axis. The apertures used to extract surface brightness profiles are shown overlaid on the radio images in Appendix~\ref{sec:azimuth_apers}.
\par
Fig.~\ref{fig:galaxy_radio_profiles} shows a separate panel for each frequency band and within each panel the surface brightness profiles are coloured according to the difference in azimuthal angle relative to the tail direction ($\Delta \theta$). The tail direction is taken to be directly east, and thus for our convention, $\Delta \theta = 0\deg$ corresponds to the tail direction and $\Delta \theta = 180\deg$ corresponds to west (i.e.\ opposite to the tail direction). The sharp decline in flux over the western edge of the disc noted by-eye in Fig.~\ref{fig:radio_imgs_highres} is confirmed quantitatively in Fig.~\ref{fig:galaxy_radio_profiles}. Namely, the steep decline of the yellow and orange lines relative to the black and purple lines. This is clearest in the $144$, $400$, $600$, and $700\,\mathrm{MHz}$ images, and is likely due to ram-pressure driven compression and truncation of the ISM within the disc. There is also a small increase in surface brightness just past $5\,\mathrm{kpc}$ prior to the steep decline. This is either due to the increased star formation along the leading edge of the disc (as noted by \citealt{tomicic2018}), increased radio emission due to the compression of magnetic field lines, or a combination of both. Discriminating between these possibilities will be addressed in a future paper in this series.
%\par
%Given the radio continuum images presented in this section, specifically the ${\sim}100\,\mathrm{kpc}$ one-sided tail to the east and the compressed radio contours and sharp surface brightness decline at the western edge of the disc, it is now more clear than ever that NGC 2276 is being strongly affected by ram-pressure stripping as it traverses in central regions of the NGC 2300 group. While there is surely some tidal interaction between NGC 2276 and NGC 2300 (e.g. the extended stellar distribution for NGC 2300, \citealt{tomicic2018}), tidal stripping tends to give rise to symmetric tidal tails \textbf{(REF)} which are not seen in these data. Therefore tidal effects are almost certainly sub-dominant to ram pressure in setting the observed morphology of NGC 2276 and its tail (see also \citealt{rasmussen2006,wolter2015}).

\section{Spectral Ageing and Plasma Speed} \label{sec:spec_age}

\begin{figure*}
    \centering
    \includegraphics[width = 0.95\textwidth]{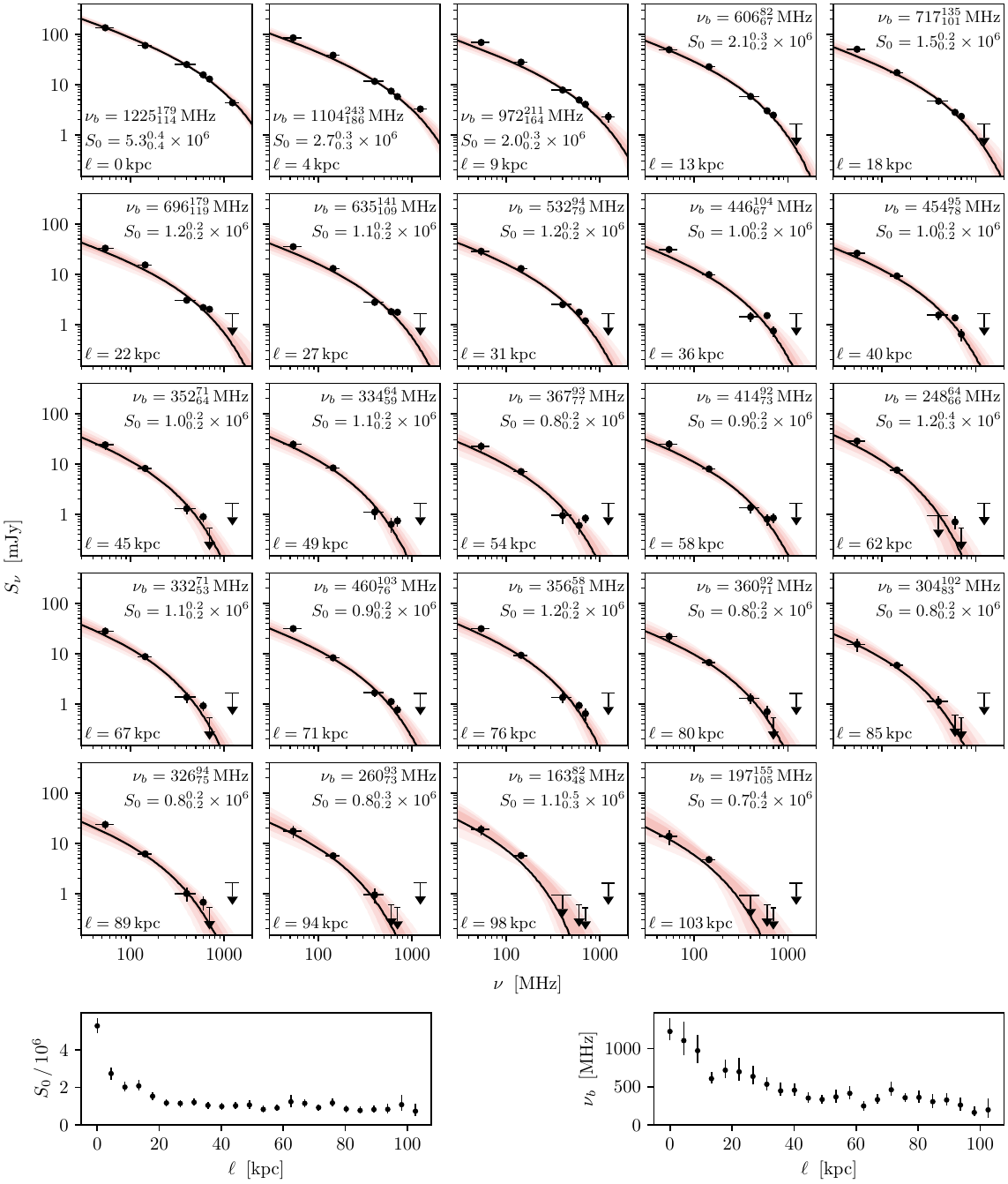}
    \caption{Radio continuum spectral-fitting results. \textit{Top:} Radio continuum spectra as a function of distance along the stripped tail. Spectra are extracted from the apertures shown in Fig~\ref{fig:spec_rad}, and distance along the tail increases in each panel from left-to-right and top-to-bottom. Measured flux densities are shown with the black data points and the best-fit spectral model is shown by the solid black line. Red shading shows the 68, 95, and 99.7 per cent model credible regions. \textit{Bottom:} Best-fit JP model parameters (normalization and break frequency) as a function of distance along the tail.}
    \label{fig:spec_rad}
\end{figure*}

In this section we explore synchrotron ageing in more detail by extracting radio continuum spectra along the tail with our full suite of continuum imaging. We then fit these spectra with a simple ageing model in order to estimate plasma ages and the stripped plasma bulk velocity.

\subsection{Measurement of the Radio Continuum Spectra} \label{sec:measure_spec}

We measured flux densities at six frequencies along the stripped tail of NGC 2276. Flux densities were measured within rectangular apertures where the height is set to encompass the north-south extent of the main tail and the width is set to $25\arcsec$, just larger than the beam major axis for the LOFAR LBA image. The set of apertures used is shown in Fig.~\ref{fig:radio_imgs} (bottom left).
\par
For each frequency band, flux densities were summed within each aperture. We estimated the noise-based uncertainty ($\sigma_\mathrm{noise}$) on the aperture flux ($S_\nu^\mathrm{ap}$) by measuring `sky' fluxes, within equally-sized apertures, from blank regions around the galaxy and tail. We then calculated $\sigma_\mathrm{noise}$ as the standard deviation of these sky flux measurements. We consider an aperture to have detected radio flux if $S_\nu^\mathrm{ap} / \sigma_\mathrm{noise} \ge 3$. When quoting our final aperture flux-density values we also include a contribution from the calibration uncertainty in the error bar. The relative calibration uncertainties ($\delta_\mathrm{cal}$) that we assume for each band are listed in Table~\ref{tab:sensitivity_table}, and the total flux-density uncertainty is calculated as
\begin{equation}
    \sigma_\mathrm{tot} = \sqrt{\sigma_\mathrm{noise}^2 + \left(S_\nu^\mathrm{ap}\,\delta_\mathrm{cal}\right)^2}.
\end{equation}
\noindent
For apertures with non-detections we quote upper limits equal to $3\sigma_\mathrm{tot}$. We measured aperture flux densities at the native resolution for each image as listed in Table~\ref{tab:sensitivity_table}. We also tested measuring flux densities for all frequencies at a matched beam size of $25\arcsec \times 25\arcsec$ and found that this did not alter any of the conclusions from this work.
\par
Measured flux densities as a function of position along the tail and frequency are listed in Table~\ref{tab:flux_table}. We note that for all frequency bands we have assumed that the radio continuum emission is entirely due to non-thermal, synchrotron radiation with no contribution from thermal, free-free emission. This is most relevant at $1.2\,\mathrm{GHz}$ where free-free emission can make a non-zero contribution to the radio continuum flux density. Typically, the `thermal fraction' at ${\sim}1\,\mathrm{GHz}$ is $\lesssim\!10\%$ \citep[e.g.][]{condon1992,niklas1997,tabatabaei2017,ignesti2022_gasp}, therefore this is likely a reasonable approximation even for our highest frequency band.

\begin{table*}[!ht]
\centering
\caption{Tail flux densities}
\label{tab:flux_table}
\begin{tabular}{lcccccc}
\toprule
\toprule
$\ell\,\mathrm{[kpc]}$ & \multicolumn{6}{c}{$S_\nu\,\mathrm{[mJy]}$} \\
\cmidrule{2-7}
 & $54\,\mathrm{MHz}$ & $144\,\mathrm{MHz}$ & $400\,\mathrm{MHz}$ & $600\,\mathrm{MHz}$ & $700\,\mathrm{MHz}$ & $1230\,\mathrm{MHz}$ \\
\midrule
0 & $134.8 \pm 14.2$ & $59.8 \pm 6.0$ & $25.1 \pm 1.3$ & $15.6 \pm 0.8$ & $12.9 \pm 0.7$ & $4.4 \pm 0.6$ \\
4 & $85.2 \pm 9.6$ & $38.4 \pm 3.9$ & $11.6 \pm 0.7$ & $7.4 \pm 0.4$ & $5.8 \pm 0.4$ & $3.3 \pm 0.6$ \\
9 & $68.8 \pm 8.1$ & $27.9 \pm 2.8$ & $7.8 \pm 0.5$ & $4.9 \pm 0.3$ & $4.1 \pm 0.3$ & $2.3 \pm 0.6$ \\
13 & $49.3 \pm 6.6$ & $22.6 \pm 2.3$ & $5.8 \pm 0.4$ & $3.0 \pm 0.3$ & $2.5 \pm 0.2$ & $<1.5$ \\
18 & $50.4 \pm 6.6$ & $17.3 \pm 1.8$ & $4.7 \pm 0.4$ & $2.8 \pm 0.2$ & $2.3 \pm 0.2$ & $<1.5$ \\
22 & $32.9 \pm 5.4$ & $15.3 \pm 1.6$ & $3.1 \pm 0.3$ & $2.2 \pm 0.2$ & $2.0 \pm 0.2$ & $<1.5$ \\
27 & $35.3 \pm 5.6$ & $13.0 \pm 1.4$ & $2.8 \pm 0.3$ & $1.8 \pm 0.2$ & $1.8 \pm 0.2$ & $<1.5$ \\
31 & $28.7 \pm 5.2$ & $13.0 \pm 1.4$ & $2.5 \pm 0.3$ & $1.8 \pm 0.2$ & $1.2 \pm 0.2$ & $<1.5$ \\
36 & $31.1 \pm 5.3$ & $9.8 \pm 1.1$ & $1.4 \pm 0.3$ & $1.5 \pm 0.2$ & $0.7 \pm 0.2$ & $<1.5$ \\
40 & $26.2 \pm 5.1$ & $9.3 \pm 1.1$ & $1.5 \pm 0.3$ & $1.4 \pm 0.2$ & $0.6 \pm 0.2$ & $<1.5$ \\
45 & $24.0 \pm 4.9$ & $8.2 \pm 1.0$ & $1.3 \pm 0.3$ & $0.9 \pm 0.2$ & $0.0 \pm 0.2$ & $<1.5$ \\
49 & $24.6 \pm 5.0$ & $8.3 \pm 1.0$ & $1.1 \pm 0.3$ & $0.6 \pm 0.2$ & $0.7 \pm 0.2$ & $<1.5$ \\
54 & $22.4 \pm 4.9$ & $7.1 \pm 0.9$ & $0.9 \pm 0.3$ & $0.6 \pm 0.2$ & $0.8 \pm 0.2$ & $<1.5$ \\
58 & $24.8 \pm 5.0$ & $8.0 \pm 0.9$ & $1.4 \pm 0.3$ & $0.8 \pm 0.2$ & $0.8 \pm 0.2$ & $<1.5$ \\
62 & $28.5 \pm 5.2$ & $7.6 \pm 0.9$ & $<0.9$ & $0.7 \pm 0.2$ & $<0.6$ & $<1.5$ \\
67 & $28.0 \pm 5.2$ & $8.7 \pm 1.0$ & $1.4 \pm 0.3$ & $0.9 \pm 0.2$ & $<0.6$ & $<1.5$ \\
71 & $31.5 \pm 5.4$ & $8.3 \pm 1.0$ & $1.7 \pm 0.3$ & $1.1 \pm 0.2$ & $0.8 \pm 0.2$ & $<1.5$ \\
76 & $31.4 \pm 5.3$ & $9.3 \pm 1.1$ & $1.3 \pm 0.3$ & $0.9 \pm 0.2$ & $0.6 \pm 0.2$ & $<1.5$ \\
80 & $21.9 \pm 4.8$ & $6.7 \pm 0.8$ & $1.3 \pm 0.3$ & $0.7 \pm 0.2$ & $<0.6$ & $<1.5$ \\
85 & $15.3 \pm 4.6$ & $5.9 \pm 0.8$ & $1.1 \pm 0.3$ & $<0.6$ & $<0.6$ & $<1.5$ \\
89 & $23.7 \pm 4.9$ & $6.2 \pm 0.8$ & $1.0 \pm 0.3$ & $0.7 \pm 0.2$ & $<0.6$ & $<1.5$ \\
94 & $17.5 \pm 4.7$ & $5.7 \pm 0.8$ & $0.9 \pm 0.3$ & $<0.6$ & $<0.6$ & $<1.5$ \\
98 & $18.9 \pm 4.7$ & $5.7 \pm 0.8$ & $<0.9$ & $<0.6$ & $<0.6$ & $<1.5$ \\
103 & $13.9 \pm 4.5$ & $4.8 \pm 0.7$ & $<0.9$ & $<0.6$ & $<0.6$ & $<1.5$ \\
\bottomrule
\end{tabular}
\end{table*}

\subsection{Spectral Model} \label{sec:spec_model}

Given multi-frequency flux densities measured along the stripped tail, we now define a spectral model which we fit to the synchrotron spectra in order to constrain plasma ages as a function of distance from the galaxy disc. For each distance aperture we fit the low-frequency radio spectrum with a \citet{jaffe1973} model (`JP model') describing a population of cosmic-ray electrons subject to synchrotron and inverse-Compton losses. The JP model comes with three free parameters: the dimensionless spectral normalization ($S_0$), the injection spectral index ($\alpha_0$), and the location of the exponential break in the spectrum ($\nu_b$). We assume a fixed value for the injection spectral index of $\alpha_0 = -0.6$, which is consistent with the spectral index at low-frequency within the stellar disc of NGC 2276 (see Fig.~\ref{fig:spec_maps_tail}). The spectral normalization and break frequency are then determined by fitting a JP model to the observed spectrum for each distance bin. This approach is similar to the ageing models used by \citet{ignesti2023_a2255} and \citet{roberts2024_radiospec_coma}, though given the larger number of frequency bands and higher image quality in this work, we are able to use a model with fewer constraints, \textit{a priori}.
\par
In practice, we follow \citet{roberts2024_radiospec_coma} and implement the JP models with the \texttt{synchrofit} Python package \citep{turner2018a,turner2018b} in order to compute theoretical synchrotron spectra given values for $S_0$, $\alpha_0$ (fixed at $-0.6$), and $\nu_b$. The fitting is performed using the \texttt{emcee} Markov-chain Monte Carlo package \citep{foreman-mackey2013} to ensure that we have robust measurements of the uncertainties on the best-fit parameters. For cases where one or more of the frequency bands have a flux-density non-detection, we include $3\sigma$ upper limits in our likelihood function following \citet{sawicki2012}.

\subsection{Spectral Fitting Results and Orbital Speed} \label{sec:fit_results}

\begin{figure}
    \centering
    \includegraphics[width = \columnwidth]{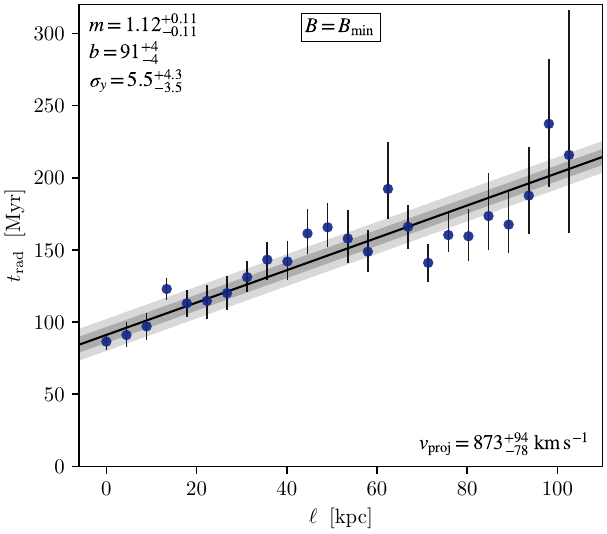}
    \caption{Radiative age versus offset from the galaxy disc for the plasma in the stripped tail. Solid line shows the best-fit linear relationship ($t_\mathrm{rad} = m \times \ell + b$) and the shaded regions show $1\times$ and $2\times$ the best-fit scatter. We also list the projected plasma velocity determined from the slope of the relation. Radiative ages are estimated assuming a minimum-loss magnetic field strength.}
    \label{fig:age_profile}
\end{figure}

In Fig.~\ref{fig:spec_rad} we show the radio continuum spectra as a function of distance along the tail, $\ell$. We note that $\ell = 0$ does not correspond to the galaxy centre, but to just off of the stellar disc to the east (see Fig.~\ref{fig:radio_imgs}). The distance along the tail increases for each panel from left-to-right and top-to-bottom. The best-fit JP model spectrum is shown in each panel with the solid black line and the red shading corresponds to the 68\%, 95\%, and 99.7\% credible regions. In the bottom row of Fig.~\ref{fig:spec_rad} we also directly show the best-fit normalizations and break frequencies as a function of the distance along the tail.
\par
The JP model provides an excellent fit to the data for every aperture. In each panel we list the best-fit normalization and break frequency. Outside of the first few apertures, the best-fit normalization remains constant at $1-2 \times 10^6$ across all distances. This is consistent with a relatively steady-state framework where the rate of plasma being removed from the galaxy disc and transported along the tail is constant (at least over the cosmic ray lifetimes that we are sensitive to at these frequencies, $\sim$hundreds of megayears).
\par
NGC 2276 clearly displays an ongoing starburst on the western edge of the disc (Fig.~\ref{fig:radio_imgs}, e.g.\ \citealt{shakhbazyan1973,tomicic2018}), thought to be induced by ram pressure. It is possible that the source of the large spectral normalization just off of the galaxy disc is due to increased production of cosmic-ray electrons from this increased star formation. Below we convert the derived break frequencies from Fig.~\ref{fig:spec_rad} to radiative ages of the plasma. Fig.~\ref{fig:age_profile} shows that the typical age for the apertures with the smallest offset from the galaxy is ${\sim}100\,\mathrm{Myr}$, which is on the same order as the expected timescale associated with the star formation visible in the near-UV image in Fig.~\ref{fig:radio_imgs}. Another possibility is that there is a gradient in the magnetic field strength, such that it is higher near the galaxy disc.
\par
As expected given the spectral index maps in Fig~\ref{fig:spec_maps_tail}, the best-fit break frequencies broadly move to lower frequency as the distance along the tail increases.  This introduces curvature and steepening to the spectra. While the decrease in $\nu_b$ with distance is not formally monotonic, given the measurement uncertainties, the data are still well described by a model of steadily decreasing $\nu_b$. Quantitatively, the relationship between $\ell$ and $\nu_b$ has a Spearman correlation coefficient of $-0.90$, indicating a strong, negative correlation.
\par
Given a value for the magnetic field strength in the stripped tail, the break frequencies determined in Fig.~\ref{fig:spec_rad} can be converted to radiative ages for the plasma using the following relation \citep{miley1980}
\begin{equation} \label{eq:trad}
    t_\mathrm{rad} \simeq 3.2 \times 10^{4}\,\frac{\sqrt{B}}{B^2 + B_\mathrm{CMB}^2}\frac{1}{\sqrt{\nu_b (1 + z)}}\;\mathrm{Myr},
\end{equation}
where $z$ is the source redshift, $B$ is the magnetic field strength in $\mathrm{\mu G}$, and $B_\mathrm{CMB}$ is cosmic microwave background equivalent magnetic field given by $B_\mathrm{CMB} = 3.25 (1 + z)^2\;\mathrm{\mu G}$. In Fig.~\ref{fig:age_profile} we show the derived radiative ages as a function of distance along the tail. Given that there are no constraints on the magnetic field strength in the NGC 2276 tail, we calculate radiative ages assuming a minimum-loss magnetic field strength ($B_\mathrm{min} = B_\mathrm{CMB} / \sqrt{3} \simeq 1.9\,\mathrm{\mu G}$) but note that this is the major uncertainty in this part of the analysis (see \citealt{ignesti2023_a2255, roberts2024_radiospec_coma} for a more detailed discussion).
\par
In Fig.~\ref{fig:age_profile} we also show the best-fit linear relationship between $t_\mathrm{rad}$ and $\ell$. The (inverse) slope of this relationship gives the best-fit velocity of the stripped plasma across the plane of the sky. For our best-fit slope, assuming a minimum-loss magnetic field, this gives a projected velocity of $874_{-78}^{+93}\,\mathrm{km\,s^{-1}}$. Given the errorbars, the data in Fig.~\ref{fig:age_profile} are well described by a linear relationship, and there is no suggestion that a quadratic relationship would provide a better fit. This indicates that cosmic-ray electrons are transported along the tail through advection and not diffusion, in agreement with previous works \citep{murphy2009_virgo,vollmer2013,muller2021,ignesti2022_gasp,roberts2024_radiospec_coma}.

\section{Discussion} \label{sec:discussion}

\begin{figure}
    \centering
    \includegraphics[width = \columnwidth]{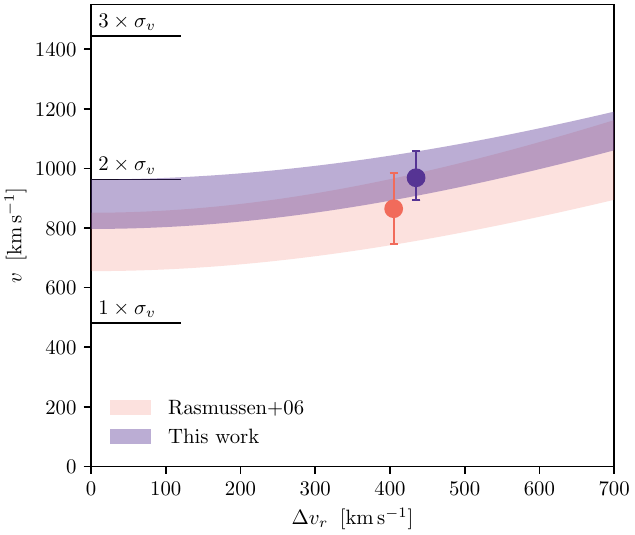}
    \caption{Total galaxy velocity as a function of, assumed, line-of-sight galaxy speed. The result from this work is shown in purple, which is compared to the galaxy speed for NGC 2276 derived by \citet{rasmussen2006} (in red). We show both our `best' estimates of the total velocity assuming $\Delta v_r = 420\,\mathrm{km\,s^{-1}}$ \citep{rasmussen2006} (data points) as well as the $1\sigma$ band for a range of assumed $\Delta v_r$ (shaded regions). The two data points are artificially offset from $\Delta v_r = 420\,\mathrm{km\,s^{-1}}$ to improve readability. One, two, and three times the group velocity dispersion \citep{finoguenov2006} are marked, assuming that $\sigma_v = \sqrt{3} \times \sigma_\mathrm{1D}$.}
    \label{fig:vel_comparison}
\end{figure}

\subsection{Constraints on the Orbital Speed of NGC 2276} \label{sec:orbital_speed}

The plasma ageing analysis in Sect.~\ref{sec:spec_age} constrains the speed of the stripped tail along the plane of the sky. Typically this tangential component ($v_t$) is the more difficult to constrain observationally, as line-of-sight velocities are relatively easily measured through Doppler shifting. Thus measuring ages along stripped tails from radio-continuum data provides a unique opportunity to estimate full, three-dimensional orbital speeds in concert with line-of-sight velocity estimates.
\par
In Sect.~\ref{sec:spec_age} we estimate a projected velocity across the plane of the sky for the stripped tail of $v_t = 874_{-78}^{+93}\,\mathrm{km\,s^{-1}}$. We make the simplifying assumption that this velocity for the stripped tail is equal to the velocity of the galaxy across the sky (see \citealt{ignesti2023_a2255} for a more detailed discussion of this point). We do not precisely know the line-of-sight velocity component for NGC 2276; however, we can approximate it as the difference in radial velocity between NGC 2276 and NGC 2300 (the central galaxy in the group). This assumes that NGC 2300 is at rest relative to the centre of the group potential. We note that the diffuse X-ray peak is spatially coincident with the location of NGC 2300 on the sky, and is roughly $6\arcmin$ to the southeast of NGC 2276. The radial velocity difference is $\Delta v_r \simeq 420\,\mathrm{km\,s^{-1}}$ \citep{rasmussen2006}, therefore our best-estimate for the three-dimensional velocity for NGC 2276 is $v = 968_{-78}^{+93}\,\mathrm{km\,s^{-1}}$.
\par
To gauge the plausability of this value, consider that the radial velocity dispersion for the NGC 2300 group is $\sigma_\mathrm{1D} \simeq 278\,\mathrm{km\,s^{-1}}$ \citep{finoguenov2006}. Assuming isotropy, this gives a three-dimensional velocity dispersion of $\sigma_v = \sqrt{3}\,\sigma_\mathrm{1D} \simeq 480\,\mathrm{km\,s^{-1}}$, and thus our best-estimate velocity for NGC 2276 would be offset from the group centroid by ${\sim}2 \times \sigma_v$. This is within the range of plausibility, especially given that NGC 2276 has a small (projected) radial offset from the group centre of ${\sim}60\,\mathrm{kpc} \simeq 0.18\,\mathrm{R_{500}}$, which allows relatively large velocity offsets while still remaining within the gravitationally-bound region of phase space.
\par
\citet{rasmussen2006} also estimate a value for the orbital velocity of NGC 2276, using a completely independent method to the analysis in this work. \citeauthor{rasmussen2006} find evidence for a mild shock front on the western edge of NGC 2276 with \textit{Chandra} X-ray data. Their estimated Mach number of $\mathcal{M} \simeq 1.7$ and sound speed of $c_s \simeq 500\,\mathrm{km\,s^{-1}}$ translate to a shock (and thus galaxy) speed of $865 \pm 120\,\mathrm{km\,s^{-1}}$. Taken at face value, the tangential velocity that we derive from the radio continuum spectra confirms that NGC 2276 is travelling at a supersonic speed since $v \ge v_t > c_s$.
\par
The velocity estimate from \citet{rasmussen2006} is remarkably similar to the estimate that we derive, indeed the two are equal within uncertainties. This supports the fact that NGC 2276 is orbiting with a velocity of ${\sim}900\,\mathrm{km\,s^{-1}}$. We re-stress that these two velocity estimates are derived from independent methodologies. In Fig.~\ref{fig:vel_comparison} we show a graphical summary of this discussion. The best-estimate velocities for NGC 2276 (assuming $v_r = 420\,\mathrm{km\,s^{-1}}$) are shown with the data points for this work (purple) and \citet{rasmussen2006} (red). Given that there is still some uncertainty around the true radial velocity for NGC 2276, relative to the group potential, we also show predictions from this work and \citet{rasmussen2006} for a range of assumed $\Delta v_r$. Regardless of the assumed $\Delta v_r$ (over a reasonable range), the two estimates agree well.

\subsection{Magnetic Field Strength in the Tail} \label{sec:Bfield}

The dominant uncertainty on the derived plasma velocity, and subsequently the galaxy orbital velocity, is the magnetic field strength in the tail. Given the lack of constraints we assume a minimum-loss magnetic field strength (${\sim} 1.9\,\mathrm{\mu G}$ for NGC 2276), but it is certainly possible that the true magnetic field strength differs from this value\footnote{We also assume that the field strength is constant along the tail, which may or may not be true in reality.}. By using the minimum-loss magnetic field strength, the plasma velocity that we estimate in Sect.~\ref{sec:fit_results} is formally a lower limit. Given Eq.~\ref{eq:trad} for the radiative age of the plasma, we can instead parameterize our estimated plasma velocity as
\begin{equation} \label{eq:v_scaling}
    v_t \simeq 874\,\mathrm{km\,s^{-1}}\,\left(\frac{N^2 + 3}{4\sqrt{N}}\right),
\end{equation}
\noindent
where $N = B / B_\mathrm{min}$.  If one takes the derived plasma velocity as reliable, then Eq.~\ref{eq:v_scaling} can be used alongside the group velocity dispersion ($\sigma_v$) to derive some rough constraints on the average magnetic field strength in the tail. This relies on the fact that a velocity greater than $3 \times \sigma_v$ is unlikely for a bound group member \citep[e.g.][]{rines2003,serra2011}, therefore $N$ in Eq.~\ref{eq:v_scaling} can only be so large.

\begin{figure}
    \centering
    \includegraphics[width = \columnwidth]{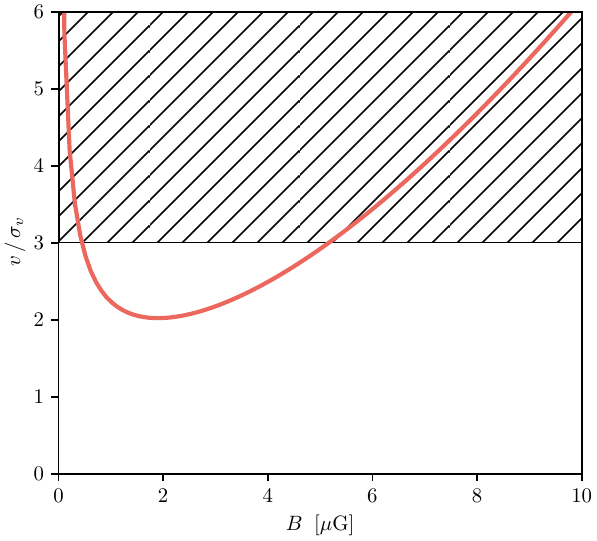}
    \caption{Normalized orbital velocity as a function of magnetic field strength, as derived from Eq.~\ref{eq:v_scaling} (red line). The hatched region corresponds to $v / \sigma_v > 3$, extreme velocity offsets that are not likely for bound satellite galaxies in groups/clusters.}
    \label{fig:B_constraints}
\end{figure}

In Fig.~\ref{fig:B_constraints} we plot the normalized three-dimensional velocity ($v / \sigma_v$, again assuming $\Delta v_r = 420\,\mathrm{km\,s^{-1}}$) as a function of magnetic field strength as given by Eq.~\ref{eq:v_scaling}. Requiring $v / \sigma_v < 3$ implies $0.5 \lesssim B \lesssim 5.0\,\mathrm{\mu G}$. This is broadly in line with the small number of prior observational constraints on the magnetic field strength in ram pressure tails \citep{muller2021,vollmer2021,ignesti2022_meerkat}.
\par
One can also estimate an equipartition magnetic field strength, though this approach is hampered by significant geometric uncertainties relating to the three-dimensional shape and filling factor of the emitting plasma. We use Eqs. (25) and (26) from \citet{govoni2004}, with $k = \Phi = \alpha = 1$ and assuming a cylindrical geometry. This gives an equipartition field strength of $B_\mathrm{eq} \simeq 1.5\,\mathrm{\mu G}$, though again we note that uncertainties on $k$ (the ratio of the energy in relativistic protons to electrons) and $\Phi$ (the volume filling factor) are substantial.
\par
Here we have derived some rough constraints on the average magnetic field strength in the tail through indirect measures, the results of which suggest that the field strength is likely less than $5\,\mathrm{\mu G}$. Stronger constraints could be derived moving forward by measuring polarized flux within the tail \citep{muller2021}.

\subsection{The Extreme Stripped Tail} \label{sec:tail_discussion}

This work makes clear the existence of a $100\,\mathrm{kpc}$ ram-pressure stripped tail behind NGC 2276, though the question of why such an extreme tail is associated to a galaxy in such a low-mass group still remains. At $144\,\mathrm{MHz}$, the NGC 2276 tail is longer than all ram-pressure tails previously imaged by LOFAR \citep{roberts2021_LOFARclust,roberts2021_LOFARgrp,ignesti2022_gasp,ignesti2022_meerkat,roberts2022_perseus,ignesti2023_a2255,hu2024}, including those for galaxies in massive clusters. The longest tail from the \citet{roberts2021_LOFARclust} sample is NGC 4848 with a projected tail length of ${\sim}40-50\,\mathrm{kpc}$ measured from the edge of the stellar disc (CGCG 97-073 and 97-079 in Abell 1367 have similar tail lengths, though both are embedded in the northwest radio relic making it difficult to measure a reliable tail size).  At a distance of ${\sim}100\,\mathrm{Mpc}$, Coma is roughly a factor of three more distant than NGC 2276, thus for an equal noise level (in Jy) we are sensitive to a fainter radio luminosity by nearly a factor of ten for NGC 2276 compared to NGC 4848. That said, extrapolating the exponential tail profile for NGC 4848 \citep{roberts2024_radiospec_coma} fainter by a factor of ten only extends the tail length by ${\sim}20\,\mathrm{kpc}$. Therefore the outlier tail length for NGC 2276 does not seem to be solely due to its vicinity, though this may still be a contributing factor.
\par
The high SFR of NGC 2276 may play a role, in the sense that this would imply a large population of cosmic-ray electrons available to be stripped from the disc. Though the SFR for NGC 2276 is certainly high, it is not unprecedented (within $2-3\sigma$ of the star-forming main sequence, e.g.\ \citealt{speagle2014,chang2015}). Tying back to the previous paragraph, NGC 4848 also has an SFR of ${\sim}10\,\mathrm{M_\odot\,yr^{-1}}$ (derived from optical+IR SED fitting, \citealt{salim2016,salim2018}), also with an ongoing starburst along the leading edge. We note that NGC 2276 and NGC 4848 also have very similar stellar masses \citep{salim2016,salim2018,tomicic2018}.
\par
It is possible that we are simply catching NGC 2276 at an opportune moment prior to the diffuse outer parts of the tail mixing with the intra-group/cluster medium. Given the direction of the stripped tail and the location of the diffuse X-ray peak of the NGC 2300 group \citep{mulchaey1993,finoguenov2006}, NGC 2276 is likely at (or very near) peak ram pressure. The less extreme intra-group/cluster medium conditions in a low-mass group compared to a massive cluster may allow such a tail to persist longer prior to mixing, though this still does not explain why no tails of comparable length (even within a factor of two) were detected in the group survey from \citet{roberts2021_LOFARgrp}.
\par
Lastly, the general picture for these non-thermal ram-pressure tails is that frozen-in magnetic fields are stripped along with the ISM, thereby allowing for extra-planar synchrotron radiation. If this scenario is correct, then we also would expect a similar tail tracing the thermal ISM. A short (${\sim}10\,\mathrm{kpc}$) X-ray tail has been detected \citep{rasmussen2006}, but no where near the extent of the low-frequency radio continuum tail (though those observations are limited by the sensitivity of current X-ray facilities to faint and diffuse soft X-ray emission). Previous \textsc{Hi} observations do not show much emission beyond the stellar disc \citep{davis1997}. In a future work we will further probe any potential \textsc{Hi} tail with our uGMRT band five data, though given our noise level and limited sensitivity to $\gtrsim\!5-10\arcmin$ scales, we do not expect to detect an \textsc{Hi} tail on the same scale as the radio continuum tail. NGC 2276 is too far north to be observable by MeerKAT, but deep VLA observations with compact configurations (C or D) could in principle detect an extended $\textsc{Hi}$ tail. Even then the LAS ($16\arcmin$ for C- and D-configuration\footnote{\url{https://science.nrao.edu/facilities/vla/docs/manuals/oss/performance/resolution}}) is comparable to the source extent in the radio continuum ($10\arcmin$). A number of ram-pressure stripped tails are $\mathrm{H\alpha}$ bright (maybe all given deep enough observations), which provides another approach for detecting a corresponding thermal tail. This could be probed photometrically using narrowband imaging from a host of northern observatories. It is also well suited to the large FOV of the large IFU mode of the WEAVE spectrograph on the William Herschel Telescope. The $\mathrm{H\alpha}$ surface brightness of ram pressure tails in low-$z$ cluster galaxies are on the order of $21 - 26\,\mathrm{mag\,arcsec^{-2}}$ \citep{yagi2010, boselli2018}. This sensitivity can be achieved by a number of northern telescopes with observations on the order of hours (or less for the bright end). Moving forward, these avenues for detecting a thermal component to the stripped tail will be pursued.

\section{Conclusion} \label{sec:conclusion}

In this work we have presented new multi-frequency radio continuum imaging of the nearby group galaxy, NGC 2276. With these data we give new constraints on the impact of ram-pressure stripping and environmental quenching in the NGC 2300 galaxy group. The main results from this work are the following:
\begin{itemize}
    \itemsep0.5em
    \item We show that NGC 2276 hosts a $100\,\mathrm{kpc}$ radio continuum tail at low frequencies. This is a factor 10 larger than previous measurements of the tail size. To our knowledge this represents the longest ram-pressure stripped radio continuum tail ever observed.

    \item With sub-kiloparsec imaging of the galaxy, we show that there is a sharp drop-off in radio continuum surface brightness along the western edge of the disc. This is consistent with compression from ram pressure.

    \item Spectral index maps and radio continuum spectra extracted along the tail are consistent with a framework where cosmic rays are removed from the disc via ram pressure and then age as they move along the tail due to synchrotron and inverse-Compton losses.

    \item We estimate the plasma velocity across the plane of the sky to be $874_{-78}^{+93}\,\mathrm{km\,s^{-1}}$ and the three-dimensional orbital velocity for NGC 2276 to be $968_{-78}^{+93}\,\mathrm{km\,s^{-1}}$.  This orbital velocity estimate shows excellent agreement with previous work from \citet{rasmussen2006} while using an independent methodology.
\end{itemize}

\noindent
Subsequent papers in this series will explore the connection between radio continuum emission and star formation within the disc, constrain cosmic-ray transport in this extreme system, as well as search for extra-planar star formation in the stripped tail. The data presented in this work, combined with existing and forthcoming UV (\textit{HST, UVIT/Astrosat}), optical (\textit{HST}, PMAS/PPAK IFU), IR (\textit{Spitzer}, SOFIA), \textsc{Hi} (VLA, uGMRT), and CO/dense gas (NOEMA) data of the galaxy disc and/or tail, form an extremely rich dataset with which to constrain strong impacts of environmental perturbations in such a sparse group.

\begin{acknowledgements}
IDR acknowledges support from the Banting Fellowship Program. RJvW acknowledges support from the ERC Starting Grant ClusterWeb 804208. FdG acknowledges the support of the ERC CoG grant number 101086378. AB acknowledges financial support from the European Union - Next Generation EU. AI acknowledges the European Research Council (ERC) programme (grant agreement No. 833824, PI B. Poggianti), and the INAF founding program `Ricerca Fondamentale 2022' (project `Exploring the physics of ram pressure stripping in galaxy clusters with Chandra and LOFAR', PI A. Ignesti). This work would not have been possible without the following software packages: \texttt{AstroPy} \citep{astropy2013}, \texttt{CASA} \citep{casa2022}, \texttt{ChainConsumer} \citep{hinton2016}, \texttt{CMasher} \citep{vandervelden2020}, \texttt{DS9} \citep{ds9_2003}, \texttt{Emcee} \citep{foreman-mackey2013}, \texttt{Matplotlib} \citep{hunter2007}, \texttt{NumPy} \citep{harris2020}, \texttt{Photutils} \citep{bradley2022}, \texttt{Regions} \citep{bradley2022_regions}, \texttt{SciPy} \citep{virtanen2020}, \texttt{Synchrofit} \citep{turner2018a,turner2018b}, \texttt{WSCLEAN} \citep{offringa2014,offringa2017}. We thank the staff of the GMRT that made these observations possible. The GMRT is run by the National Centre for Radio Astrophysics of the Tata Institute of Fundamental Research. The National Radio Astronomy Observatory is a facility of the National Science Foundation operated under cooperative agreement by Associated Universities, Inc. LOFAR \citep{vanhaarlem2013} is the Low Frequency Array designed and constructed by ASTRON. It has observing, data processing, and data storage facilities in several countries, which are owned by various parties (each with their own funding sources), and which are collectively operated by the ILT foundation under a joint scientific policy. The ILT resources have benefitted from the following recent major
funding sources: CNRS-INSU, Observatoire de Paris and Université d’Orléans, France; BMBF, MIWF-NRW, MPG, Germany; Science Foundation Ireland (SFI), Department of Business, Enterprise and Innovation (DBEI), Ireland; NWO, The Netherlands; The Science and Technology Facilities Council, UK; Ministry of Science and Higher Education, Poland.
\end{acknowledgements}

\bibliographystyle{aa}
\bibliography{main}

\appendix

\section{Galaxy Azimuthal Apertures} \label{sec:azimuth_apers}

\begin{figure}
    \centering
    \includegraphics[width = \columnwidth]{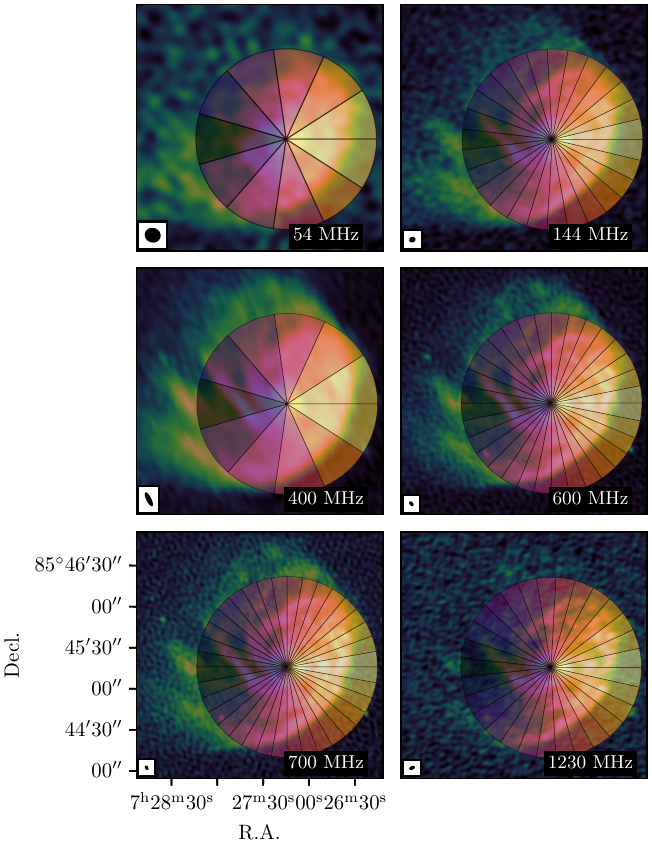}
    \caption{Same as Fig.~\ref{fig:radio_imgs_highres} but with the wedge apertures used for the surface-brightness profiles in Fig.~\ref{fig:galaxy_radio_profiles} overlaid. The radius of the wedges corresponds to the isophotal size, $R_{25}$, for NGC 2276.}
    \label{fig:azimuthal_apertures}
\end{figure}

In Fig.~\ref{fig:azimuthal_apertures} we show the wedge apertures that are described in Sect.~\ref{sec:high-res}. These apertures are used to extract surface-brightness profiles as a function of azimuthal angle from the images of NGC 2276 at different frequencies. The colouring of the apertures in Fig.~\ref{fig:azimuthal_apertures} corresponds to the colouring of the surface brightness profiles in Fig.~\ref{fig:galaxy_radio_profiles}.

\end{document}